\documentclass[aps,prfluids]{revtex4-2}
\usepackage{amssymb}
\usepackage{float}
\usepackage{multirow}
\usepackage{adjustbox}
\usepackage{graphicx}
\usepackage{color}
\usepackage[normalem]{ulem}

\begin{document}
\title{Drag anisotropy of cylindrical solids in fluid-saturated granular beds}

\author{Ankush Pal and Arshad Kudrolli}
\address{Department of Physics, Clark University, Worcester, MA 01610, USA}

\date{\today}

\begin{abstract}
We study the direction-dependent drag acting on a cylindrical solid intruder with length $L$ and diameter $D$ as it moves in water-saturated granular beds at constant depth. Polysterene and hydrogel grains with diameter $d$ are used to investigate materials which have high contact friction and those which are nearly frictionless, respectively. The drag on the intruder  is measured while oriented perpendicular $F_\perp$ and parallel $F_\parallel$ to its axis as a function of speed $U$ from the quasi-static to the rate-dependent regime.  We find that the drag anisotropy $\xi = F_\perp/F_\parallel$ is not constant, and increases significantly with driving rate and $L/D$ in both mediums. In particular for $L/D = 40$, $\xi$ increases from 2.6 to 4.5, and from 7.0 to 8.2 in the high and low friction beds, respectively, as the nondimensional Froude number $Fr 
=U/\sqrt{g(D+d)}$ is varied between $10^{-4}$ and $2 \times 10^{-2}$. On average, $\xi$ is observed to increase logarithmically with $L/D$ for $L/D \gg 1$. Exploiting the index-matched nature of hydrogel grains in water, we further show that the sediment flow around the cylinder in the two orientations is consistent with skin friction dominated drag.  The relative contributions of the cylindrical side and the circular flat-ends on $\xi$ are estimated with thin disks to understand the observed variation of drag with aspect ratio and surface friction. 
\end{abstract}

\maketitle

\section{Introduction}
The drag experienced by an intruder moving in a granular bed is fundamental to probing the unsteady rheology of granular materials and important for engineering mixers, diggers, and ploughs across a wide range of industries from mining to food processing~\cite{Albert2001,costantino2011low, balmforth2014yielding,Segiun2011}. 
The direction-dependent drag experienced by a cylindrical solid in a granular bed is important to understanding anchoring and drag-assisted propulsion of organisms and robots in sandy terrestrial and submarine environments~\cite{Jung2010,Hosoi2015,Jung2017,kudrolli2019burrowing}. 

The drag acting on an intruder depends on a number of factors including its shape and material properties, and those of the medium. 
For a circular cylinder with a length $L$ much greater than its diameter $D$, the drag while moving with speed $U$ perpendicular and parallel to its axis are given in the Stokes regime by~\cite{lamb11,batchelor1970,Yang2017}: 
\begin{equation}
F_\perp = - \frac{4 \pi \eta L}{\ln(L/D) + \gamma_\perp} U,   
\label{eq:d_per}
\end{equation}
and
\begin{equation}
F_\parallel = - \frac{2 \pi \eta L}{\ln(L/D) + \gamma_\parallel} U,   
\label{eq:d_para}
\end{equation}
respectively, where, $\eta$ is the viscosity of the medium, and $\gamma_\parallel \approx -0.114$ and $\gamma_\perp \approx 0.866$ are constants which further depend weakly on the aspect ratio $A_r = L/D$. Thus, while the drag is proportional to the relative speed, the drag anisotropy given by the ratio of the drag while moving perpendicular to its axis $F_\perp$, and the drag while moving  parallel to its axis $F_\parallel$, i.e. 
\begin{equation}
\xi = \frac{F_\perp}{F_\parallel},    
\end{equation}
is approximately $2$ in the case of long thin rods, independent of speed. The fact that $\xi \neq 1$ is important for example to swimming, because it enables a net propulsive force while using cyclic non-reciprocal strokes of the body, with a greater difference leading to faster calculated locomotion speeds~\cite{Lighthill1969,Lauga2009}. 

Granular sediments display a yield stress and shear thinning rheology different from a Newtonian fluid, and $\xi$ is therefore expected to be different. The drag experienced by spherical and cylindrical intruders has been observed to scale linearly with the overburden weight of the sediments and crosssectional area when the intruder size is much greater than the sediment size~\cite{Brzinski2010,Gulliard2014,Takada2019}. The dynamic force resulting from granular collisions becomes important with speed and has been calculated in dry granular materials~\cite{Kubota2021}. Presence of interstitial liquid influences the drag in granular beds~\citep{panaitescu2012nucleation,allen2019effective,hossain2020drag}, but remains relatively less studied compared to the dry granular case~\cite{andreotti_forterre_pouliquen_2013,Winter2014}. Various $\xi$ have been reported in studies with granular beds composed of dry glass beads including $\xi \approx 3$ in experiments with rods with $L/D \approx 2.5$ after subtracting end effects~\cite{maladen2009undulatory,Jung2010,maladen2011mechanical}. Whereas, $\xi$ ranging between 5 and 7 were measured for long thin rods moving slowly through granular hydrogels sedimented in water motivated by burrowing and biolocomotion~\citep{kudrolli2019burrowing}.  
Thus,  it is difficult to have an integrated picture from these reports because of their different focus, and a systematic study examining the issue of drag anisotropy itself in granular mediums is still necessary. 

In this paper, we report systematic measurements of drag and drag anisotropy experienced by cylindrical solids moving in two complementary fluid-saturated sediment beds with high and low grain-grain contact friction. The drag from the essentially quasistatic regime observed at low speeds, to the rate-dependent friction regime at higher speeds is analyzed. The drag anisotropy is measured as a function of aspect ratio of the cylinders and compared across the two kinds of mediums.  We analyze the relative contribution of the drag acting on the end-faces of the cylinders and the shear drag on the cylindrical surface of the solid by contrasting with measured drag in the thin disk limit. We visualize the fluorescence tagged grains around the cylinder in the case of the transparent hydrogel medium and illustrate the nature of the flow in the two orientations which gives rise to the observed drag.  
     
\section{Experimental System} 
\begin{figure*}
\begin{center}
\includegraphics[width=15cm]{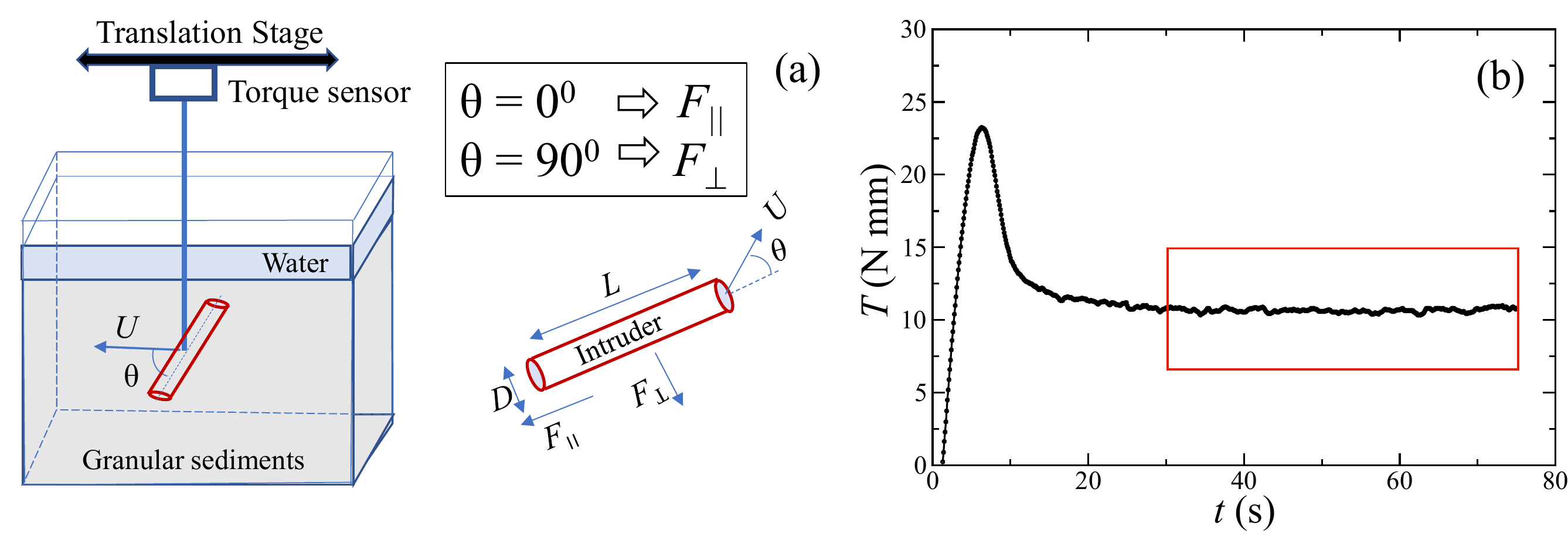}
\end{center}
\caption{(a) A schematic diagram of a cylindrical intruder dragged at constant depth inside a rectangular tank filled with granular sediments immersed in water. {\color{black} The diagram shows the intruder moving while oriented perpendicular to its axis ($\theta = 90^0$) to measure $F_\perp$. Complementary measurements are also performed  while the intruder moves parallel to its axis ($\theta = 0^0$)  to measure the drag $F_\parallel$.}  (b) Measured torque $T$ as a function of time $t$ for a cylinder dragged through water-saturated polystyrene bed ($A_r = 5$; $z_p = 4$\,cm; $U = 1$\,mm/s). The axis of the cylinder is perpendicular to the direction of motion. The drag is calculated by averaging the torque measured over the time interval indicated by the red box, once steady state is established. \label{fig:intruder}}
\end{figure*}

\begin{table}
\centering
\begin{tabular}{ c c c } \hline
Length $L$ (cm) & Diameter $D$ (cm) & Aspect Ratio $A_r$\\ \hline
\hline
8 & 1.6 & 5 \\ 
8 & 0.8 & 10 \\ 
16 & 0.8 & 20 \\ 
16 & 0.4 & 40 \\  \hline
\end{tabular}
\caption{The length $L$, diameter $D$, and aspect ratio $A_r = L/D$ of the cylindrical intruders used to measure drag. Measurements were also performed with a set of thin disks ($L = 3$\,mm) with the same set of diameters to estimate the relative contributions of the circular ends.}
\label{tab1}
\end{table}

A schematic of the experimental system used to measure drag is shown in Fig.~\ref{fig:intruder}(a). Cylindrical solids composed of clear resin were fabricated using a 3D printer (Formlabs Form 3). Intruders with dimensions listed in Table~\ref{tab1} are used to vary aspect ratio $A_r = L/D$. The intruder is attached rigidly to a thin steel arm, which is connected to a torque sensor (Mark 10 Model 5i) placed on a translating stage with a computer controlled stepper motor (Zaber). The intruder can be thus moved with a prescribed speed $U$ ranging between 0.05\,mm/s and 8\,mm/s. Then, the flow inertia relative  to the gravitational field is given by  the Froude Number ${\rm Fr} = \frac{U}{\sqrt{g(D+d)}}$~\cite{Chehata2003}, where $g = 9.8$\,m/s$^2$, ranges between $10^{-4}$ and $2 \times 10^{-2}$ over the system parameters investigated. 
The rod is immersed in the granular bed with its major axis oriented perpendicular or parallel to the direction of motion as it moves horizontally at constant depth as shown in Fig.~\ref{fig:intruder}(a).  

Spherical polystyrene grains with mass density $\rho_{g} =$ 1050\,kg/m$^{3}$, mean diameter $d = 0.45$\,mm, and granular hydrogels with density $\rho_{g}$= 1005 kg/m$^{3}$ and mean diameter $d = 1$\,mm immersed in distilled water with density $\rho_w = 997$\,kg/m$^{3}$, viscosity $\eta_f= 1$\,mPa\,s are used in the study. The polystyrene and hydrogel grains have a coefficient of friction $\mu_s \approx 0.5$, and $10^{-3}$~\cite{brodu2015spanning}, respectively, spanning the range typically observed in granular mediums ranging from sand to tapioca. The medium is placed inside a rectangular tank with a length $L_c = 32$\,cm, width $W_c = 22$\,cm, and height $H_c =  22$\,cm, and settles to the tank bottom because $\rho_{g} > \rho_w$. A granular volume fraction $\phi \approx 0.6$ is observed in both kinds of sedimented beds consistent with previous observations~\cite{panaitescu2012nucleation,allen2019effective}.  The depth of the bed is maintained at 17\,cm in both cases, with a 4\,cm water layer above to eliminate any capillary effects. The drag of the water layer is observed to be negligible compared to the sediment drag. For consistency, the bed was initialized before each trial by stirring and allowing the grains to settle over approximately 10 minutes for the polystyrene grains, and 13 minutes for the granular hydrogels. The lab temperature is controlled and maintained at $25^{\circ}$C $\pm\, 2^{\circ}$C.

Figure~\ref{fig:intruder}(b) shows a measured torque $T$ over time $t$ once the translating stage is prescribed to move. The torque increases rapidly before reaching a maximum, corresponding to the yield stress of the medium. The torque then falls and becomes nearly constant as the intruder moves with the prescribed speed. This torque combines both the drag due to the intruder and the thin arm. Therefore the torque due to the arm is measured separately, and subtracted from the total torque. The drag is obtained by dividing the resulting torque due to the intruder by the length of the arm. 

Preliminary measurements were performed at various depths $z_p$ and drag was observed to scale linearly with depth provided $z_p \gtrsim 2D$. This is consistent with previous studies which found that drag in fluid-saturated sediment beds scale with the overburden pressure $P_{p} = \phi \left(\rho_{g} - \rho_{w} \right) g z_{p}$, where $g$ is gravitational acceleration, in the quasi-static regime~\cite{costantino2011low}, and the rate-dependent regime as well~\cite{panaitescu2017drag,allen2019effective}.  Thus, we chose $z_p = 4$\,cm in the case of the polystyrene grains, and $z_p = 12$\,cm in the case of the hydrogel beads, to optimize for the sensitivity range of the torque sensor and to reduce the number of experimental parameters to be explored. At this depth, the drag was measured while the rods were away from the container walls in the front by $z_p$ to avoid boundary effects which were otherwise observed to lead {\color{black} to systematically higher drag in the case of the high-friction grains~\cite{allen2019effective}.} We estimated the inertial number $I = \frac{U d}{ D \sqrt{P_p/\rho_g}}$~\cite{Cruz2005,panaitescu2017drag} which measures relative importance of inertia of grains to imposed forces, and viscous number $J = \frac{\eta_f U}{P_p D}$~\cite{Boyer2011, allen2019effective} which measures the viscous effects of the interstitial fluid versus imposed forces. We find $2 \times 10^{-5} < I < 2 \times 10^{-2}$, and $1 \times 10^{-7} < J < 2 \times 10^{-4}$ in the two mediums. Thus, grain inertia and viscous forces are small compared with imposed forces in the experiments.

\section{Drag Measurements}
\begin{figure*}
\centering
\includegraphics[width=14cm]{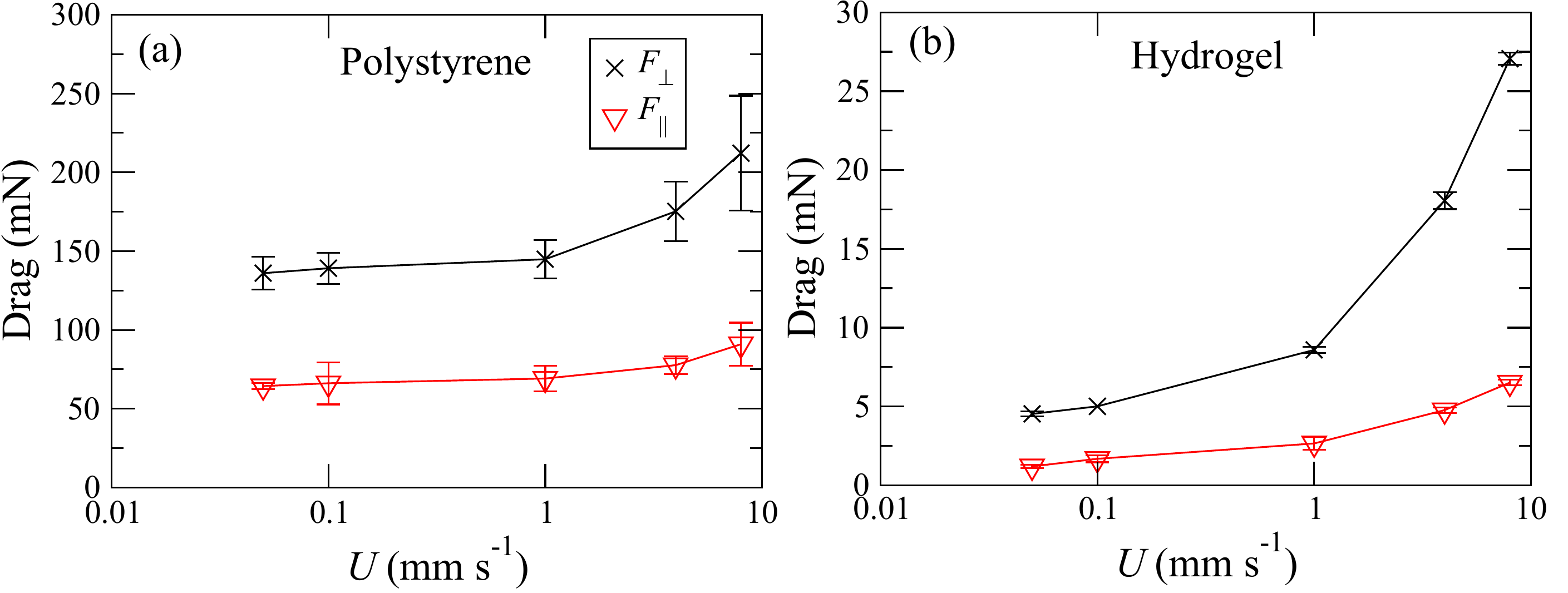}
\caption{(a-d) The measured drag $F_\perp$ and $F_\parallel$ for a cylindrical rod with $A_r = 5$ in water-saturated (a) polystyrene and (b) hydrogel granular beds as a function of speed $U$. The drag approaches a non-zero value at low speeds and increases with speeds in each case.  The errors bars correspond to RMS fluctuations observed over 6 to 12 trials each.  \label{fig:drag}}
\end{figure*}

\begin{figure*}
\centering
\includegraphics[width=13cm]{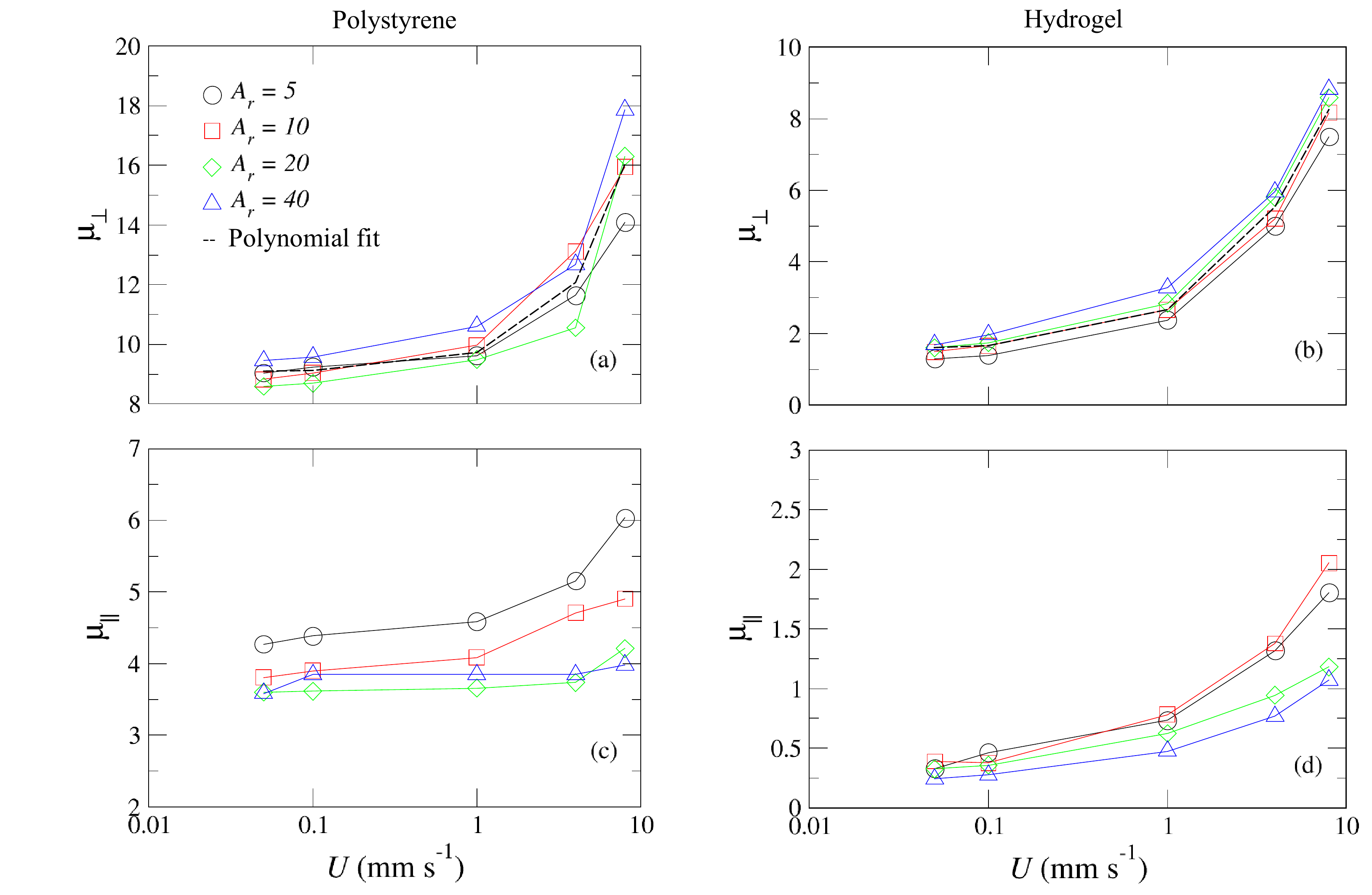}
\caption{(a-d) The effective friction as a function of speed is plotted for different aspect ratio of the cylinder in two mediums while oriented perpendicular (a,b), and parallel (c,d) to its major axis. {\color{black} A second-order polynomial fit given by Eq.~(\ref{eq:quad}) is also shown in (a) and (b). The goodness of fit is 0.99 in both cases. The linear  term is observed to be at least an order of magnitude smaller compared to the quadratic term in both mediums (see text).} \label{fig:muU}}

\end{figure*}
\begin{figure*}
\centering
\includegraphics[width=14cm]{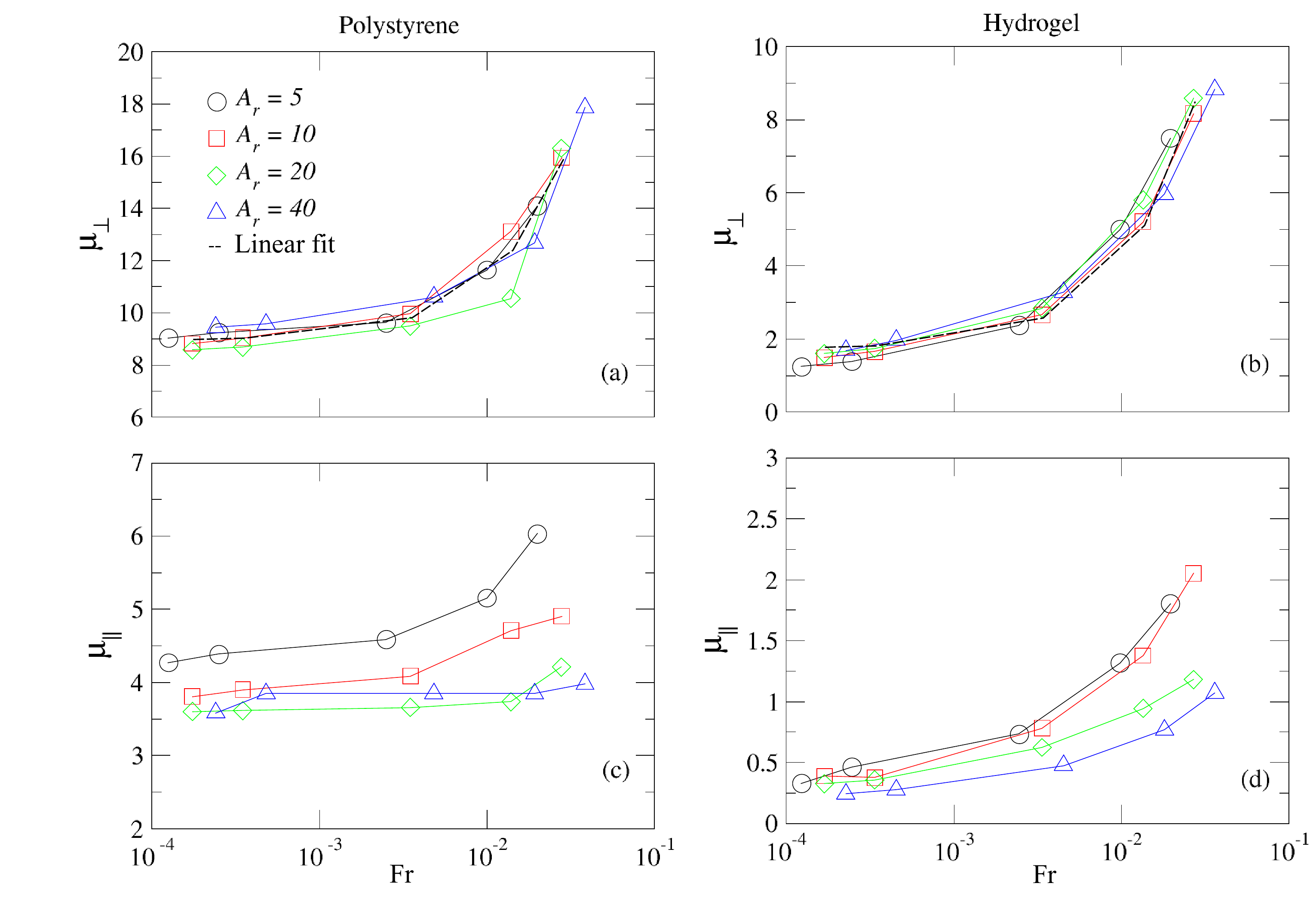}
\caption{(a-d) The effective friction in granular polystyrene beds and hydrogel beds as a function of Froude Number ${\rm Fr} = \frac{U}{\sqrt{g(D+d)}}$ while oriented perpendicular in beds composed of (a,b), and while oriented parallel (c,d). {\color{black} The data is described by a linear fit to $\mu_\perp = \mu_o + \eta \, {\rm Fr}$, where $\mu_o = 8.9$ and 1.7, $\eta = 242.5$ and 246.5, in the case of polystyrene, and hydrogel beds, respectively. The goodness of fit is 0.99 in both cases.} \label{fig:muIJ}}
\end{figure*}

Figure~\ref{fig:drag} shows $F_\perp$ and $F_\parallel$ moving in the two mediums corresponding to a cylinder with aspect ratio $A_r = 5$ over the wide range of speeds studied. Each measurement of drag corresponds to an average over at least 6 to 12 trails, and the error bars noted correspond to the root mean square (RMS) deviation from the mean. We observe that $F_\perp$ is much greater than $F_\parallel$ over the entire range of speeds, and approach non-zero values at vanishing speeds consistent with yield stress materials in both low and high friction grains~\cite{andreotti_forterre_pouliquen_2013}. Further, $F_\perp$ and $F_\parallel$ increases with speed showing a rate-dependent regime where viscous and collisional stresses can be expected to become important. In order to estimate the relative contribution of skin friction drag and pressure drag, we estimate the force due to pressure difference as $F_p = \frac{1}{2}\rho U^2 LD$, and find that the ratios $F_{\perp,\parallel}/F_p \gg 10^2 $ and 
$10^3$, in the case of polysterene and hydrogel grains, respectively. Thus, it appears that the friction drag dominates, and that the pressure drag is unimportant over the range of speeds in our study. {\color{black} Pressure drag is typically dominant when there is separation of boundary layer rear of the intruder. However, the visualization of flow (see Fig.~\ref{fig:flow}) around the cylinder does not reveal any vortex or wake, supporting the fact that pressure drag is not significant in our system.}

To cast the measured drag in nondimensional form, we obtain the effective friction as the ratio of the drag, and the weight of the granular bed given by the overburden pressure $P_p$ acting on the projected area of the cylinder $LD$.  Then,  
\begin{equation}
\mu_{\bot} = \frac{F_{ \bot}}{P_p DL},
\end{equation}
and
\begin{equation}
\mu_{\|} = \frac{F_{ \|}}{P_p DL}.
\end{equation}
With these definitions, we have $\xi = \mu_{\bot}/\mu_{\|}$ consistent with $\xi = F_{\bot}/F_{ \|}$.

Figure~\ref{fig:muU} shows a plot of the measured effective friction as a function of speed in both orientations for the two mediums. In the quasi-static limit, we observe  $\mu_{\bot} \sim 9$ in the case of the polystyrene grains which have high grain surface friction. A significantly lower $\mu_{\bot} \sim 1.5$ is observed in the hydrogel grains which are nearly frictionless. An effective friction $\mu_{\bot} \sim 13$ has been reported in the quasi-static limit of a steel rod moving in the perpendicular orientation in dry glass beads which have a grain surface friction $\mu_s$ in the range 0.5 and 0.7, more comparable to polystyrene~\cite{Gulliard2014}. Further, Fig.~\ref{fig:muU}(a,b) shows that $\mu_{\bot}$ increases roughly together with increasing speed for the intruders with various $A_r$ in both mediums. While Fig.~\ref{fig:muU}(c,d) shows that $\mu_{\|}$ also increases with speed, it is systematically lower compared with $\mu_{\bot}$. However, it can be observed that $\mu_{\|}$ does not collapse, but is lower at higher $A_r$. Thus, the effective friction encountered by the intruder can be rate dependent besides varying with its aspect ratio, and grain-grain friction properties of the medium. 

{\color{black} We fit $\mu_\perp$ with a second-order polynomial given by 
\begin{equation}
    \mu_\perp = \mu_o + \gamma_1 U + \gamma_2 U^2,
    \label{eq:quad}
\end{equation}
where $\mu_o$ corresponds to the yield-stress, $\gamma_1$ and $\gamma_2$ are fitting constants corresponding to the linear and quadratic terms in $U$. From the fits to Eq.~(\ref{eq:quad}) shown in Fig.~\ref{fig:muU}(a,b), we find $\gamma_1 = 0.63$\,s/mm and $\gamma_2 = 0.029$\,s$^2$/mm$^2$ in the case of polystyrene beds, and $\gamma_1 = 1.16$\,s/mm and $\gamma_2 = 0.04$\,s$^2$/mm$^2$ in the case of hydrogel beds. Since $\gamma_2 \ll \gamma_1$, the linear term dominates considering that $U <10$\,mm/s in our experiments. Thus, the effective friction encountered by the cylindrical intruder while moving perpendicular to its axis can be described as increasing linearly starting from a value corresponding to the yield stress over the range speeds investigated in our experiments in both mediums. }

Because various $D$ are used in the investigation of the effect of aspect ratio, we use the Froude Number ${\rm Fr}$ to nondimensionalize the speed variable, and plot $\mu_{\bot}$ and $\mu_{\|}$ versus ${\rm Fr}$ in Fig.~\ref{fig:muIJ} in the two mediums. While the collapse is arguably better in the perpendicular direction, the systematic variation in the parallel orientation remain even after accounting for the differences in the diameters. {\color{black} Because the increase in  $\mu_\perp$ could be described by the linear term in Eq.~(\ref{eq:quad}), we fit $\mu_\perp$ as a function of Fr in Fig.~\ref{fig:muIJ}(a,b) to $\mu_\perp = \mu_o + \eta \, {\rm Fr}$, where $\mu_o = 8.9$ and 1.7, $\eta = 242.5$ and 246.5, in the case of polystyrene, and hydrogel beds, respectively.} (We also plotted $\mu_{\bot}$ and $\mu_{\|}$ as a function of the viscous number $J$ and the inertial number $I$ in the case of the polysterene and hydrogel grains, respectively, but did not find any significant improvement in the collapse.)  
Thus, these plots show that while $\mu_{\bot}$ is independent of the aspect ratio over $5 \leq L/D \leq 40$, $\mu_{\|}$ decreases over the same range. This is in contrast with a Newtonian fluid in the viscous regime, where a little over 50\% reduction can be calculated in the drag per unit length in both orientations according to Eqs.~(\ref{eq:d_per}) and (\ref{eq:d_para}) over the same range of $L/D$.   

\section{Drag Anisotropy}
\begin{figure}
\centering
\includegraphics[width=14cm]{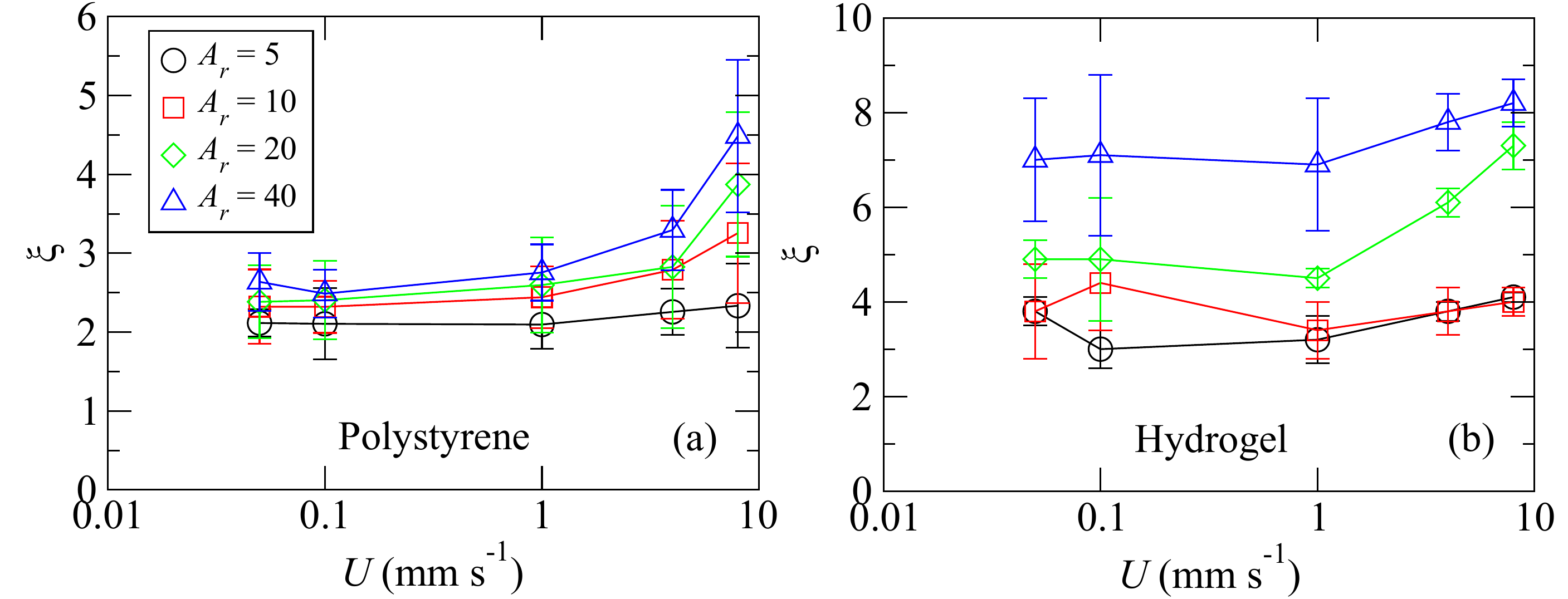}
\caption{The drag anisotropy $\xi = F_\perp/F_\parallel$ for cylinder with various aspect ratio as a function of speed in water-saturated granular (a) polystyrene and (b) hydrogel beds. The error bars correspond to RMS values. 
\label{fig:xi}}
\end{figure}

We plot $\xi = F_\perp/F_\parallel$ to analyze the degree of drag anisotropy in the two mediums as a function of speed for the various intruders in Fig.~\ref{fig:xi}, and also list them in Table~\ref{tab:p}. We observe that the overall drag anisotropy is significantly greater in the low-friction hydrogels  compared to the high-friction polystyrene. Further, we observe that $\xi$ appears to be more or less constant at lower $A_r$. Whereas, $\xi$ increases with speed as $A_r$ increases in the case of the hydrogel bed, and more so in the case of the polystyrene bed.  Thus, $\xi$ in both kinds of granular sediments are greater than that for a Newtonian fluid, and further show significant rate and aspect ratio dependence.  We understand from Fig.~\ref{fig:muU} that this dependence arises from the fact that $\mu_\|$ decreases with $A_r$, but $\mu_\bot$ remains relatively unchanged with $A_r$. The rate dependence of $\xi$ can be linked to the fact that $\mu_\|$ changes more slowly compared with $\mu_\bot$ with increasing speed.

\begingroup
\begin{table*}
\centering
Polystyrene\\

{\begin{tabular}{|c  c | c| c c | c| c c | c| c c | c|}
\hline
\multicolumn{3}{|c |}{ $A_r$ = 5} & \multicolumn{3}{|c |}{$A_r$ = 10} & \multicolumn{3}{|c |}{$A_r$ = 20} & \multicolumn{3}{|c |}{$A_r$ = 40} \\ \hline
$U$ (mm/s) & $\xi$ & $\overline{\xi}$ & $U$ (mm/s) & $\xi$ & $\overline{\xi}$ & $U$ (mm/s) & $\xi$ & $\overline{\xi}$ & $U$ (mm/s) & $\xi$ & $\overline{\xi}$ \\  \hline
0.05 & 2.1 $\pm$ 0.2 & \multirow{5}{*}{2.2} & 0.05 & 2.3 $\pm$ 0.5 & \multirow{5}{*}{2.6} & 0.05 & 2.4 $\pm$ 0.5 & \multirow{5}{*}{2.8} & 0.05 & 2.6 $\pm$ 0.3 & \multirow{5}{*}{3.1} \\  
0.1 & 2.1 $\pm$ 0.4 & & 0.1 & 2.3 $\pm$ 0.3 &  & 0.1 & 2.4 $\pm$ 0.5 &  & 0.1 & 2.5 $\pm$ 0.3 &  \\ 
1 & 2.1 $\pm$ 0.3 & &  1 & 2.4 $\pm$ 0.4 &  & 1 & 2.6 $\pm$ 0.6 & & 1 & 2.7 $\pm$ 0.3 &  \\ 
4 & 2.2 $\pm$ 0.3 & & 4 & 2.8 $\pm$ 0.6 &  & 4 & 2.8 $\pm$ 0.8 & & 4 & 3.3 $\pm$ 0.5 &  \\
8 & 2.3 $\pm$ 0.5 & & 8 & 3.2 $\pm$ 0.9 &  & 8 & 3.9 $\pm$ 0.9 & & 8 & 4.5 $\pm$ 1 &  \\ \hline
\end{tabular}}\\
\vspace{0.5cm}
Hydrogel\\

{\begin{tabular}{|c  c | c| c c | c| c c | c| c c | c|}
\hline
\multicolumn{3}{|c |}{$A_r$ = 5} & \multicolumn{3}{|c |}{$A_r$ = 10} & \multicolumn{3}{|c |}{$A_r$ = 20} & \multicolumn{3}{|c |}{$A_r$ = 40} \\ \hline
$U$ (mm/s) & $\xi$ & $\overline{\xi}$ & $U$ (mm/s) & $\xi$ & $\overline{\xi}$ & $U$ (mm/s) & $\xi$ & $\overline{\xi}$ & $U$ (mm/s) & $\xi$ & $\overline{\xi}$ \\  \hline 
0.05 & 3.8 $\pm$ 0.3 & \multirow{5}{*}{3.6} & 0.05 & 3.8 $\pm$ 1 & \multirow{5}{*}{3.9} & 0.05 & 4.9 $\pm$ 0.4 & \multirow{5}{*}{5.5} & 0.05 & 7.0 $\pm$ 1.3 & \multirow{5}{*}{7.4} \\ 
0.1 & 3.0 $\pm$ 0.4 &  & 0.1 & 4.4 $\pm$ 1 &  & 0.1 & 4.9 $\pm$ 1.3 &  & 0.1 & 7.1 $\pm$ 1.7 & \\ 
1 & 3.2 $\pm$ 0.5 & & 1 & 3.4 $\pm$ 0.6 &  & 1 & 4.5 $\pm$ 0.2 &  & 1 & 6.9 $\pm$ 1.4 & \\ 
4 & 3.8 $\pm$ 0.2 &  &  4 & 3.8 $\pm$ 0.5 &  & 4 & 6.1 $\pm$ 0.3 &  & 4 & 7.8 $\pm$ 0.6 & \\ 
8 & 4.1 $\pm$ 0.1 & & 8 & 4.0 $\pm$ 0.3 &  & 8 & 7.3 $\pm$ 0.5 & & 8 & 8.2 $\pm$ 0.5 &  \\ \hline
\end{tabular}}
\caption{The measured drag anisotropy of cylinders of various aspect ratios moving through water-saturated sedimented polystyrene and hydrogel mediums.     \label{tab:p}}
\end{table*}
\endgroup

\begin{figure}
\centering
\includegraphics[width=7cm]{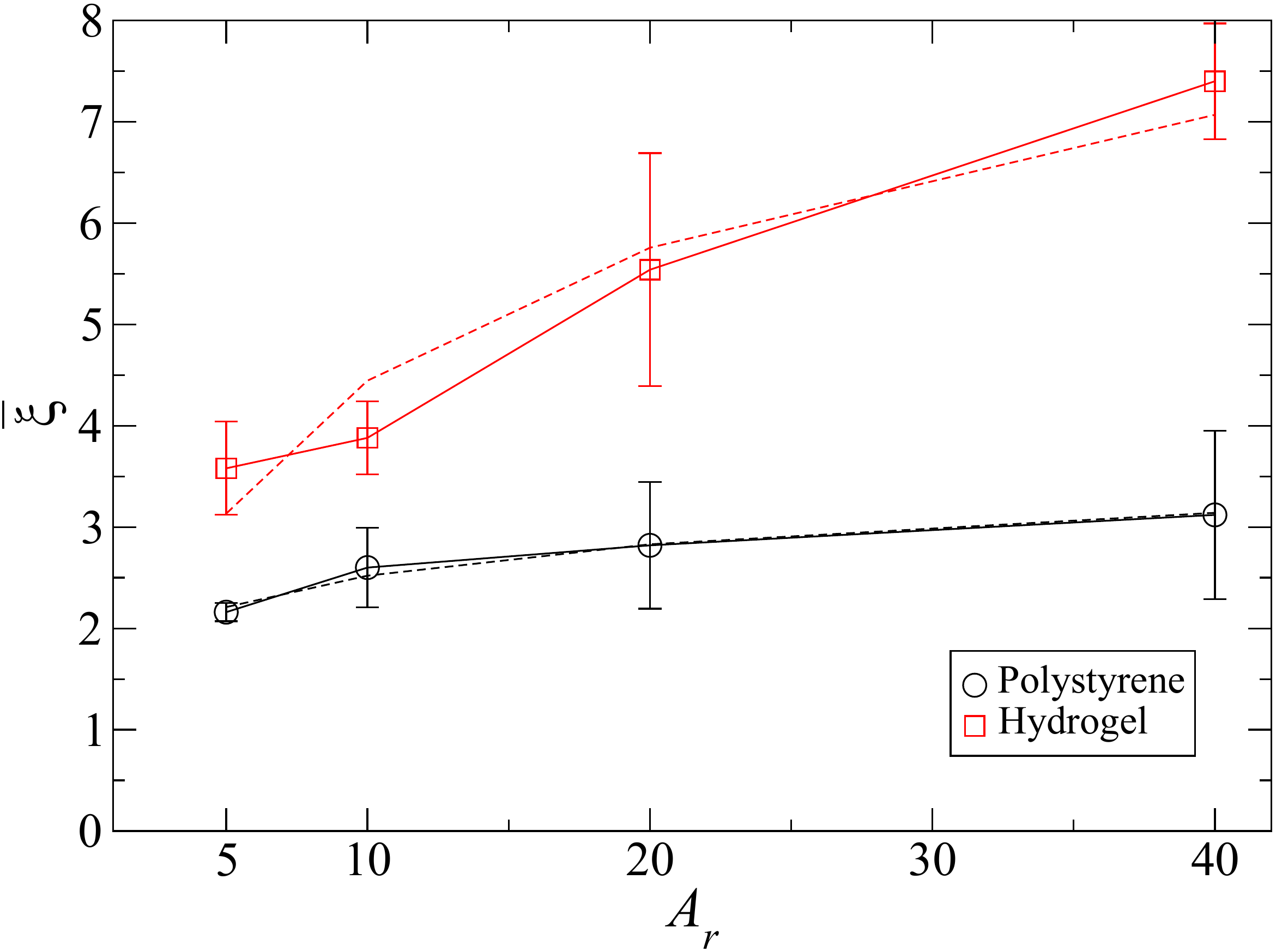}
\caption{The average drag anisotropy over speed $\overline{\xi}$ is observed to increase with $A_r$, and is greater in low-friction hydrogels compared with high-friction polystyrene grains. The data is described by the logarithm fit (dashed lines) given by Eq.~\ref{eq:xi_fit} (Polystyrene: $\alpha  = 0.44, \beta =  1.49$; Hydrogel: $\alpha = 1.89, \beta =  0.085$). \label{fig:drag_Ar}}
\end{figure}

To show the overall trends with $A_r$, we average over $\xi$ observed across $U$, and plot the resulting average drag anisotropy $\overline{\xi}$ in Fig.~\ref{fig:drag_Ar}. The minimum and maximum range of $\xi$ observed over $U$ are shown with  error bars. We clearly observe that $\overline{\xi}$ increases with $A_r$ in both mediums. While  $\overline{\xi}$ increases from about 2.2 to $3.1$ in the high-friction polystyrene, $\overline{\xi}$ increases from 3.6 to 7.4 in the low-friction medium as $A_r$ increases from 5 to 40. Noting the fact that the drag appears to scale with $L$ in the perpendicular orientation, we fit the data with the logarithmic form inspired by Eq.~(\ref{eq:d_para}), i.e. 
\begin{equation}
\overline{\xi} = \alpha \ln(L/D) + \beta 
\label{eq:xi_fit}
\end{equation}
where, $\alpha$ and $\beta$ are material dependent fitting constants. We fit $\overline{\xi}$  in Fig.~\ref{fig:drag_Ar} for $L/D \geq 5$, and find that it captures the overall trends well with a stronger dependence on the $L/D$-term in the case of the low-friction hydrogel grains. The fact that $\alpha$ is much greater in the high-friction material compared to the low-friction material is notable, and further work is needed to fully explore the connection between the implied stronger dependence and material friction.   

\section{Cylindrical side versus flat end drag}
\begin{figure}
\centering
\includegraphics[width=12cm]{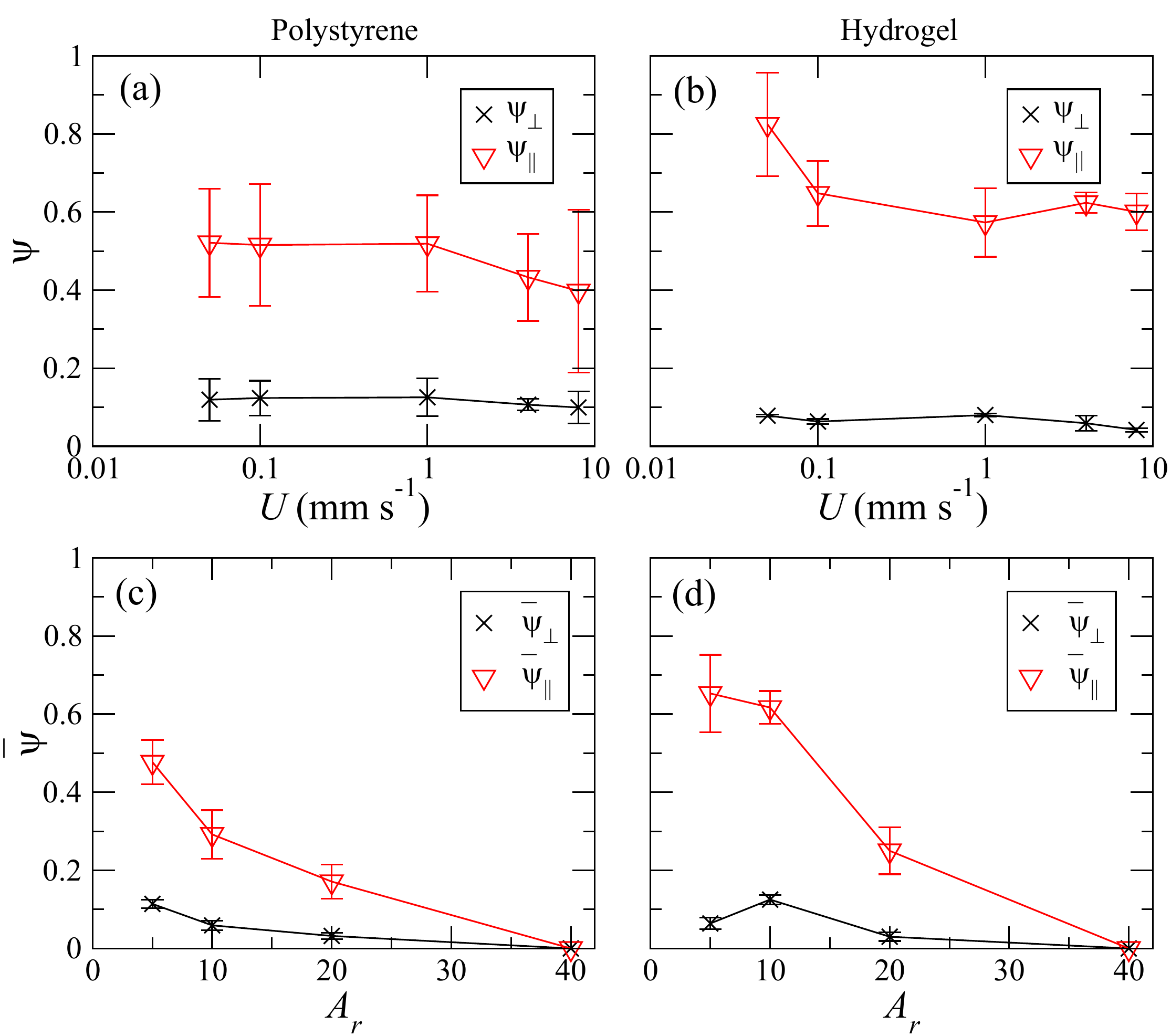}
\caption{(a,b) The fractional disk to cylinder drag $\psi_{\perp,\parallel}$ given by the ratio of  $F_{\perp,\parallel}$ corresponding to the disk and the $A_r = 5$ cylinder with same diameter as a function of speed for (a) polystyrene, and (b) hydrogels. The fractional contribution is nearly constant across intruder velocity. (c,d) The fractional disk to cylinder drag averaged over speeds $\overline{\psi}$ decreases with $A_r$ in both polystyrene (c), and hydrogels (d).  \label{fig:drag_fraction}}
\end{figure}
Besides the stresses acting on the cylindrical sides of the intruder, the drag acting on the flat circular ends is important to determining the total drag in the two orientations, and thus the drag anisotropy. In order to estimate the relative contribution of these surfaces, we manufactured 2 to 3\,mm thick circular disks with the same set of diameters as the cylindrical intruders. We measure $\overline{\xi} \approx 0.35$ and $0.53$ corresponding to $A_r = 0.3/1.6 \approx 0.2$, in the case of the hydrogel and polystyrene mediums, respectively. These values of $\overline{\xi} < 1$  may be expected when $L/D \ll 1$ because only the shear stress on the end-faces contribute to drag in the disk limit while in the perpendicular orientation, whereas the normal stress on the end-faces remained essentially unchanged while in the parallel orientation. Further, $\overline{\xi} < 1$ can be also expected, extrapolating from the fact that $\overline{\xi} \sim 1$ when $L/D \sim 1$.   

Figure~\ref{fig:drag_fraction}(a,c) shows the fraction of drag $\psi_{\perp,\parallel}$ corresponding to the ratio of drag of a thin disk with $A_r = 0.2$ and a cylinder with $A_r = 5$ across the range of $U$ investigated in the two mediums.  We find in both mediums that the $\psi_\perp$ is less than 20\% across the range of $U$ studied. Whereas $\psi_\parallel$ is about 50\% and 60\% in the polystyrene and hydrogel mediums, respectively. We further plot the fractional disk to cylinder drag averaged over speeds $\psi_\perp$ and $\psi_\parallel$ as a function of $A_r$ in Fig.~\ref{fig:drag_fraction}(c,d). These ratios are observed to decrease rapidly with $A_r$. Thus, the drag contributions of the ends in the parallel orientation are always greater and decay slower, than when in the perpendicular orientation.  As $A_r$ increases above 20, the contribution of the ends decrease to less than 10\%. By $A_r = 40$, the contributions of the cylinder ends to drag were found to be less than the sensitivity of our instruments. Since  $\mu_\bot$ continues to decrease on average at these large aspect ratios, the increase in $\xi$ appears to be linked to the continued evolution of the tangential stresses acting on the cylindrical surface of the intruder while oriented in the parallel orientation.         

Although the number of sediment grains in contact with the cylinder at a given time instant can be assumed to be approximately similar in both orientations, the number of grains which actually need to rearrange as the intruder moves through the same distance is vastly different when $L/D \gg 1$. {\color{black} The time scale over which a typical sediment grain is sheared by the intruder can be expected to scale as $D/U$ in the perpendicular orientation, and as $L/U$ in the parallel orientation.} Thus, a smaller fraction of granular sediments are sheared over a sustained longer time in the parallel orientation compared to the perpendicular orientation as the intruder moves by. The resulting difference in shear induced dilatancy and the lowering of local granular packing fraction~\cite{Mueller2010} may be the reason for the lower $\mu_\|$ compared to $\mu_\bot$, and why it continues to decrease with $A_r$. {\color{black} Skin drag exists for the cylinder moving in either orientation. The granular medium fluidized by the advancing intruder stays attached to the intruder over a longer length while orientated parallel to the direction of motion. Because a lubrication layer between grains can occur when the grains are in motion, this may result in a lower skin friction, and hence in a higher $F_\perp$ compared to $F_\parallel$.}

\section{Flow profiles}
\begin{figure*}
\centering
\includegraphics[width=17 cm]{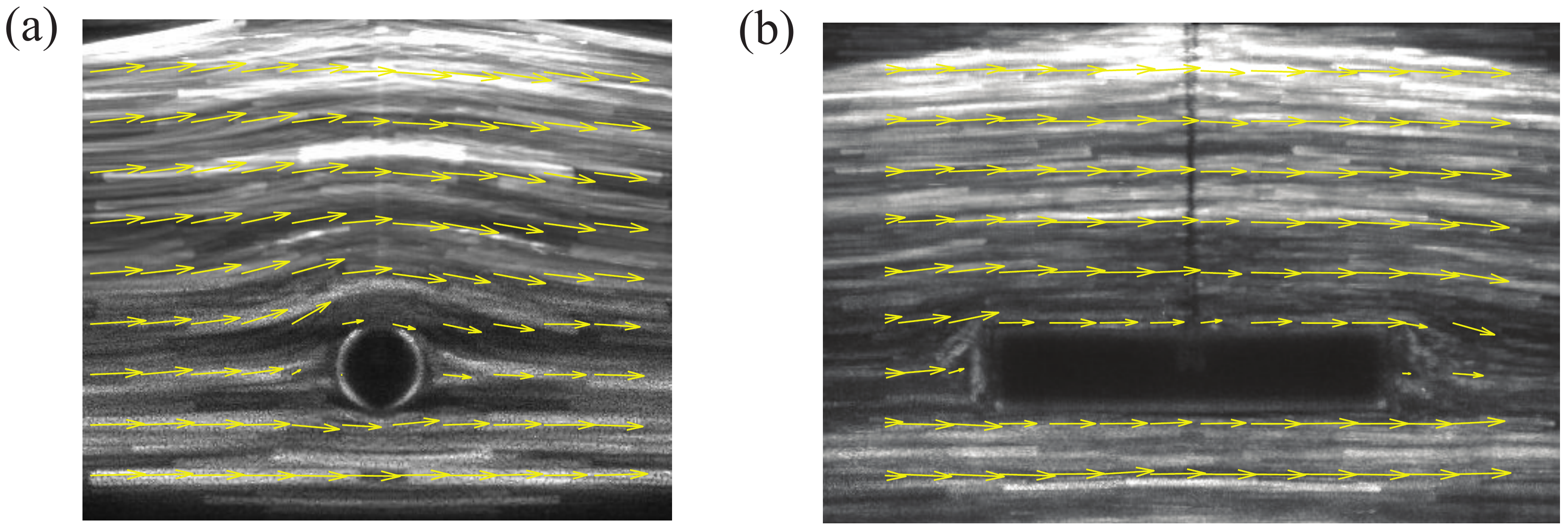}
\caption{The granular flow in the frame of reference of a cylindrical intruder moving (a) perpendicular and (b) parallel to its axis  ($D=1.6$\,cm; $L = 8$\,cm; $A_r = 5$; $U = 1$\,mm/s). The flow field is superimposed on a time integrated image obtained by tracing bright pixels over 50\,s. A central plane is illuminated using a laser sheet. The flow can be observed to be nearly symmetric about the central vertical axis. Systematic asymmetry can be observed  above and below the intruder because of gravitational symmetry breaking. 
\label{fig:flow}}
\end{figure*}

We examine the sediment flow around the cylinder while moving in the two orientations to complement the $\xi$ measurements. The flow around the cylinder moving through water-saturated hydrogels grains can be visualized exploiting the fact that their refractive index nearly matches that of water. This method enables us to measure the flow in a frame of reference in which the intruder is fixed to make the flow measurements simpler. A smaller 15\,cm cubical tank is used for this purpose and is filled with hydrogel beads to a depth of 12\,cm with a 3\,cm water layer on top. The tank is placed on a linear track which can move forward or backward with the same velocities as those used in measuring drag. The cylindrical intruder with $L/D = 5$ is attached with a fixed arm and placed inside the water saturated granular particles at a depth $z_p= 6$\,cm. About 10\% of the hydrogel beads were hydrated with water mixed with a dye (Rhodamine 6G, Eastman) to use as tracers and mixed in with the rest of the grains. A thin laser sheet (Z Laser, Germany) illuminates the fluorescent dye soaked hydrogel beads around the intruder.  A low-light 16-bit camera (C 11440, Hamamatsu) fitted with a lens and bandpass filter captures the motion of the hydrogel beads around the cylinder as the rod is moved relative to the medium. 
A sequence of images between 1.25 and 30 frames per second was captured  as $U$ was increased from 0.05\,mm/s to 8 mm/s, when the container with the medium was moved through 5\,cm.

Figure~\ref{fig:flow} shows the granular flow observed past the cylindrical intruder while oriented perpendicular and parallel to the direction of relative motion with speed 1\,mm/s. The captured images are time averaged over a 50\,second time interval using ImageJ to generate bright steaks corresponding to the fluorescent grains to illustrate the flow past the cylinder. No vortices are observed in the wake which would signal the importance of form drag over the entire range of speeds studied. Similar flow profiles are observed over the entire range of speeds studied, and appear to be consistent with those where drag is dominated by skin friction. {\color{black} This is in contrast to the observation of vortex structures behind a sphere-shaped intruder moving through similar granular sediments~\cite{jewel2018micromechanics}. However, the diameter of the intruders studied there were considerably larger. Thus, the differences in the geometry of the flow, and greater $D$ appear to imply that the greater speeds needed to typically see vortices in dense granular flows~\cite{Samadani2003,Johnson2011}, are not reached in our experiments. } 

We tracked the tracer grains by using standard software which finds the centroid associated with the bright pixels and their motion over consecutive frames~\cite{Crocker1996}. We then obtained the mean velocity over approximately  $1 {\rm cm} \times 1 {\rm cm}$ area grid points, and further time-averaged while the container and the medium are moved. These fields are also superimposed on the images in Fig.~\ref{fig:flow}. The flow appears relatively symmetric about the central vertical axis through the intruder, and approaches the imposed speed sufficiently far away from the cylinder. Whereas, differences can be observed in the relative flow above and below the intruder, with the stagnation of flow occurring below the central axis of the cylindrical rod below the center while moving in either orientation. The magnitude of the sediment field velocity was found to reach $U$ within the 20\% accuracy of our measurements at a distance of $2D$ in front and behind the intruder, and in a distance of $D$ above and below the intruder in both orientations. 

We measure the horizontal line corresponding to the stagnation points to be $\approx 0.2D$ below the cylinder axis while oriented perpendicular, and to be also $\approx 0.2D$ below the cylinder axis while oriented parallel. This symmetry breaking occurs because of gravity which results in a greater $P_p$ with increasing depth leading the flow resistance to be systematically lower above the rod. Some of these features of the flow have been noted previously in dry granular flow simulations past a cylinder oriented with its axis perpendicular to the flow in the quasi-static regime~\cite{Gulliard2014}. Thus, while we are unable to visualize the flow in the case of the opaque polystyrene medium, we expect these features of the flow illustrated in Fig.~\ref{fig:flow}, to be present there as well. 

Significant slip can be observed past the smooth sides of the intruder in Fig.~\ref{fig:flow}, similar to previous observations of granular hydrogels around spherical solid intruders~\cite{jewel2018micromechanics}, as well as in the case of the motion of glass beads moving past smooth glass surfaces which have significantly higher surface friction~\cite{Siavoshi2006}. From the nature of these observed flows, one may expect that the drag due to normal and tangential stress distributions on the intruder would continue to contribute to the drag as $L/D$ increases in the perpendicular orientation, whereas only the drag due to tangential stress distribution may be expected to remain relatively important in the parallel orientation.  We anticipate that the normal and tangential stress distribution scale differently in the two orientations from the observations that $\xi$ increase systematically with intruder speed and aspect ratio. The tangential stress given by the effective viscosity and the strain rate can be expected to change~\cite{Mueller2010,allen2019effective} as the moving intruder increasingly fluidizes the athermal granular medium with speed.   
However, further modeling work is required to connect the observed flow profiles with the measured trends in $\mu_\bot$, $\mu_\|$, and $\xi$.

\section{Conclusions}
In summary, we measured the direction-dependent drag of cylindrical intruders moving in sediment beds, and found a significant variation of drag anisotropy with aspect ratio and system material properties. The drag is examined from the quasi-static regime to the rate-dependent regime when inertial effects are unimportant. A higher drag is found in the granular sediments with higher grain-grain contact friction, but the drag anisotropy is found to be significantly greater in the lower-friction hydrogels. The drag anisotropy is observed to be greater in both mediums compared to that in a Newtonian fluid in the Stokes regime, and systematically increases with speed.  On average, drag anisotropy is observed to increase consistent with a logarithmic form as the aspect ratio of the cylinder increases in both kinds of mediums. This form of increase appears to arise due to the relative decrease of the drag per unit length in the parallel orientation, while the drag  per unit length in the perpendicular direction remains relatively unchanged for sufficiently large aspect ratios. Thus, drag and drag anisotropy of a cylindrical solid moving in granular sediments with wide ranging material properties is qualitatively different than in a viscous Newtonian fluid.

We also evaluated the contribution of the flat circular cylinder ends on the measured drag in parallel and perpendicular orientations by constructing thin disks with the same diameters. We observe that the end-faces make a greater contribution in the parallel orientation compared to the perpendicular orientation at lower aspect ratios. We visualize the flow of the granular medium around the intruders in the two orientations and show that a symmetric flow develops fore and aft of the intruder in both orientations. The flow is observed to remain attached to the intruder surface with no visible vortices in either orientations over the range of speeds investigated. 

Based on the observations on the  contributions of end-faces using the disks, and the fore-aft symmetry in the flow fields around the cylindrical intruders, it appears that skin friction dominates the drag in granular sediments in our experiments. While both normal and shear components of the flow are important all along the length of the cylinder in the perpendicular orientation, the shear component of the flow past the length of the cylinder dominates in the parallel orientation, and the flow region around the ends is relativity small. While our experiments cannot directly measure the normal and shear components of the drag in the two mediums, we believe numerical simulations can access this issue, and further illuminate the reason for the observed drag anisotropy in the two complementary mediums.

\begin{acknowledgments}   
We thank Samuel Hoang and Benjamin Allen for help with experiments and discussions, and Brian Chang for a critical reading of the manuscript and feedback. This work was supported by the U.S. National Science Foundation under Grant No. CBET-1805398.
\end{acknowledgments}   

%


\end{document}